\newcommand\be{\begin{equation}}
\newcommand{\ee}{\end{equation}}
\newcommand{\bea}{\begin{eqnarray}}\newcommand{\eea}{\end{eqnarray}}
\newcommand{\nn}{\nonumber}\newcommand{\p}[1]{(\ref{#1})}
\newcommand{\lb}[1]{\label{#1}}

\newcommand\q{\quad}

\newcommand\cE{{\cal E}}
\newcommand\cF{{\cal F}}
\newcommand\cG{{\cal G}}

\newcommand\cL{{\cal L}}

\newcommand\cP{{\cal P}}
\newcommand\cT{{\cal T}}

\newcommand\stc{\stackrel{\star}{,}}

\def\g{\gamma}

\def\d{\delta}

\def\ve{\varepsilon}

\def\bph{{\bar\phi}}

\def\k{\kappa}

\def\o{\omega}

\def\vt{\vartheta}

\def\D{\Delta}

\def\G{\Gamma}

\def\pa{\partial}

\def\na{\nabla}

\newcommand\hx{{\hat{x}}}

\newcommand\hd{{\hat\delta}}

\def\sfrac#1#2{{\textstyle\frac{#1}{#2}}}

\documentclass[12pt]{article}
\usepackage{amsmath,amssymb}

\renewcommand{\thefootnote}{\fnsymbol{footnote}}

\begin{document}
\begin{center}
{\bf  REALITY IN NONCOMMUTATIVE GRAVITY }\\

\vspace{1cm}

{\large\bf
  B.M. Zupnik\footnote{zupnik@theor.jinr.ru}}
\vspace{1cm}

{\it Bogoliubov Laboratory of Theoretical Physics, JINR, \\
141980, Dubna, Moscow Region, Russia}\\
\end{center}

\begin{abstract}
We study the problem of reality in the geometric formalism of the 4D
noncommutative gravity using the known deformation of the diffeomorphism
group induced by the twist operator with the constant deformation
parameters $\vt^{mn}$. It is shown that  real covariant derivatives
can be constructed via  $\star$-anticommutators of the real connection
with the corresponding fields. The minimal noncommutative generalization
of the real Riemann tensor contains only $\vt^{mn}$-corrections of the
even degrees in comparison with the undeformed tensor. The gauge field
$h_{mn}$ describes a gravitational field on the flat background. All
geometric objects are constructed as the perturbation series using
$\star$-polynomial decomposition in terms of $h_{mn}$. We consider
the nonminimal tensor and scalar functions of $h_{mn}$ of the odd 
degrees in $\vt^{mn}$ and remark that these pure noncommutative objects 
can be used in the noncommutative gravity.
\end{abstract}

\renewcommand{\thefootnote}{\arabic{footnote}}
\setcounter{footnote}0
\setcounter{equation}0
\section{Introduction}

The simplest noncommutativity in the 4-dimensional space is based on the
following relation for operators of coordinates $\hx^m$:
\be
\hx^m\star\hx^n-\hx^n\star\hx^m=i\vt^{mn},
\ee
where $\vt^{mn}$ are some constants and $m, n=0, 1, 2, 3$. We shall
consider the Weyl ordering in the algebra  of  noncommutative (NC) fields
$A_\vt$ using operator polynomials symmetrized in all indices
\be
\hat{\phi}(\hx^m)=f^0+f^1_m\hx^m+\sum\limits_{k=2}^\infty f^k_{(m_1
\ldots m_k)}\hx^{(m_1}\star\ldots\star\hx^{m_k)},
\ee
where $f^k_{(m_1\ldots m_k)}$ are some numerical coefficients. One can
analyze the commutative image of this operator function
\be
\hat{\phi}(\hx^m)\rightarrow \phi(x^m)=f^0+f^1_mx^m+\sum
\limits_{k=2}^\infty f^k_{(m_1\ldots m_k)}x^{m_1}\ldots x^{m_k},
\ee
We treat $\phi(x)$ as an element of the commutative algebra $A$ of smooth
functions in the space $R^4$ with coordinates $x^m$.

As it has been shown recently \cite{CKNT,We} the basic properties and
symmetries of the NC field theories in $A_\vt$ are connected with the
quantum-group structures induced by the  twist operator
\be
\cF=\exp{(-\cP)},\q \cP=\sfrac{i}2\vt^{mn}\pa_m\otimes\pa_n\lb{twist}
\ee
which acts on the tensor products of functions $\phi\otimes\chi$. In
particular, the Moyal-Weyl representation of the noncommutative product
has the following form:
\bea
&&\phi\star\chi=\mu\circ\cF^{-1}(\phi\otimes\chi)=\phi(x)\chi(x)+
\sfrac{i}2\vt^{mn}(\pa_m\phi)(x)(\pa_n\chi)(x)\nn\\
&&-\sfrac18\vt^{mn}\vt^{rs}(\pa_m\pa_r\phi)(x)(\pa_n\pa_s\chi)(x)+
O(\vt^3),\lb{star}
\eea
where $\mu(\phi\otimes\chi)=\phi\chi$ is the bilinear multiplication map
in the commutative algebra $A$. The corresponding bilinear map
in the NC algebra $A_\vt$ is defined as $\mu_\star=\mu\circ\cF^{-1}$.

The complex conjugation of this $\star$-product satisfies the relation
\be
\overline{\phi\star\chi}=\bar{\chi}\star\bph,
\ee
where $\bph$ and $\bar{\chi}$ are complex conjugated functions. The
noncommutative product of the real functions $\phi$ and $\chi$ is not a
real element of $A_\vt$
\bea
&&\phi\star\chi=\sfrac12(\phi\star\chi+\chi\star\phi)+\sfrac12(\phi\star
\chi-\chi\star\phi)\equiv\sfrac12\{\phi\stc\chi\}+\sfrac12[\phi\stc\chi],
\nn\\
&&\overline{\{\phi\stc\chi\}}=\{\phi\stc\chi\},\q\overline{[\phi\stc\chi]}
=-[\phi\stc\chi].\lb{stcomm}
\eea
Using Eq.\p{star} one can check that deformation corrections in the
$\star$-anticommutator  contain only even  degrees in $\vt$ starting from
$\vt^2$: $\{\phi\stc\chi\}= 2\phi\chi+O_+(\vt^2)$, while  the
$\star$-commutator has the odd-degree $\vt$-decomposition $[\phi\stc\chi]
=O_-(\vt)$.

The authors of refs. \cite{CKNT,We} considered the quantum-group
deformation of the Poincar\'e group using the twist operator $\cF$
\p{twist} and proved that the NC algebra $A_\vt$ is covariant with
respect to this quantum group. We will consider this twist deformation
in  Section 2.

The new approach to noncommutative gravity theory was proposed in
\cite{ABDMSW} (see, also, a  more deep discussion of the noncommutative
mathematical formalism in \cite{ADMW}). This approach is based on the
twist deformation of the diffeomorphism group in the real 4-dimensional
space. In Section 3, we consider our treatment of  the basic
principles of this approach:
1) The diffeomorphism transformations of the primary matter fields and
the metric tensor are not deformed; 2) The twisted diffeomorphism
group acts covariantly on the $\star$-products of  fields in the
noncommutative algebra. The second principle is more fundamental in
the NC gravity, while the first is convenient for the comparison with
the ordinary gravity.

In Section 4 we construct the real NC generalizations of the Christoffel
symbols and the corresponding Riemann tensor which have the standard
transformation properties in the twisted diffeomorphism group. The
condition of reality is used at all stages of our geometric formalism,
for instance, the real covariant derivative of the metric tensor vanishes
by  definition. We derive the deformed Bianchi identity for the covariant
derivatives of the NC Riemann tensor. By analogy with the ordinary
gravity, it is convenient to analyze all nonlinearities of the NC
formalism using the gravitational gauge field $h_{mn}$ on the flat-space
background. Note that the  NC geometric quantities of ref.\cite{ABDMSW}
are complex; however, they can be reduced to  real quantities plus
some complex tensors or scalars constructed from the  gravitational field.

Our `minimal' version of the real gauge invariant NC-gravity action is
constructed in Section 5. This action contains the perturbation series in
the field $h_{mn}$ starting from the standard free spin 2 term. Each
interaction term of the action is also invariant with respect to the
background twisted Poincar\'e group. Varying this action in the  field
$h_{mn}$ one obtains the real NC gravity equations transforming as the
contravariant tensor density. In our geometric formalism, the reality of
the NC deformation of the classic equations of gravity seems natural, and
the real NC field $h_{mn}$ describes a standard number of the physical
degrees of freedom. It is shown that the minimal NC-gravity equations have
no odd-degree $\vt^{mn}$-corrections in comparison with the Einstein equations.

In Section 6, we show that the formalism of noncommutative gravity is much 
more flexible than the formalism of ordinary gravity, in particular, one can
construct `nonminimal' NC tensor or scalar functions of the gravitational
field which have no classical analogues and vanish in the commutative
limit $\vt^{mn}\rightarrow 0$. The nonminimal scalars with arbitrary 
constants could be added to the NC-gravity action.  Note that the
original NC-gravity action in the complex geometric formalism is real by
definition \cite{ABDMSW}. It is not difficult to compare $h_{mn}$ 
decompositions of all versions of the NC-gravity actions. The choice of
the appropriate model of the NC gravity requires the analysis
of possible physical effects of the noncommutativity.

\setcounter{equation}0
\section{Twisted Poincar\'e symmetry}
Let us consider the infinitesimal transformations of the scalar field
$\phi(x)$ in the Poincar\'e group
\bea
&&\d \phi(x)=-(P_c+M_\o)\phi(x)=-(c^mP_m +\sfrac12\o^{mn}M_{mn})\phi(x),
\nn\\
&&P_m=\pa_m,\q M_{mn}=x_n\pa_m-x_m\pa_n,\lb{Ptrans}
\eea
where $c^m, P_m$ and $\o^{mn}, M_{mn}$ are the parameters and generators
of translations and Lorentz rotations, respectively.

The coproduct in the Poincar\'e group is trivial on the group generators
\bea
&&\D(P_c)=P_c\otimes 1+1\otimes P_c,\q\D(M_\o)=M_\o\otimes 1+1\otimes M_\o
\lb{copr}.
\eea
This coproduct determines the action of generators on the tensor and local
products of fields.

In our interpretation of the twist-deformed Poincar\'e group, the primary local 
fields of the NC theory  have  the same basic transformations 
 \p{Ptrans}. Twist deformations appear in the
coproduct of the deformed Lorentz transformations
\bea
&&\D_t(\pa_m)=\exp(-\cP)\D(\pa_m)\exp(\cP)=\D(\pa_m),\lb{coprP}\\
&&\D_t(M_\o)=\exp(-\cP)\D(M_\o)\exp(\cP)=\D(M_\o)\nn\\
&&+\sfrac{i}2\o^{mn}\vt_{ms} P_n\otimes P^s
-\sfrac{i}2\o^{mn}\vt_{rn} P^r\otimes P_m.
\lb{coprM}
\eea
This coproduct acts on the tensor product of functions
\bea
&&-\D_t(M_\o)\circ \phi\otimes\chi=(\d_\o \phi)\otimes\chi+\phi\otimes
(\d_\o \chi)\nn\\
&&+\sfrac{i}2(\o^{mn}\vt_{ns}-\vt^{mn}\o_{ns})\pa_m\phi\otimes \pa^s\chi.
\lb{dtens}
\eea
Applying the map $\mu_\star=\mu\circ\cF$ to this relation one can obtain
the covariant formula of the deformed Lorentz transformations on the
$\star$-product of the primary scalar fields
\bea
&&\hd_\o(\phi\star\chi)= -\mu_\star\circ\D_t(M_\o)\circ \phi\otimes\chi
=(\d_\o\phi)\star\chi+\phi\star(\d_\o\chi)\nn\\
&&+\sfrac{i}2(\o^{mn}\vt_{ns}-\vt^{mn}\o_{ns})\pa_m\phi\star \pa^s\phi=
-M_\o(\phi\star \chi).
\eea
The covariant deformed Lorentz transformations of $\star$-products of
tensor fields will be used in Section 5.

\setcounter{equation}0
\section{Twisted diffeomorphism group}

We shall consider the active form of the infinitesimal diffeomorphism
transformations of the real scalar field
\be
\d_\xi \phi(x)=-[\xi,\phi](x)\equiv -\cL_\xi\phi(x)=-(\xi^m\pa_m\phi)(x),
\lb{scaltr}
\ee
where $\xi^m(x)$ are arbitrary smooth functions and $\cL_\xi$ is a Lie
derivative corresponding to a differential operator $\xi=\xi^m\pa_m$.
The infinite-dimensional Lie algebra of diffeomorphisms is isomorphic to
the set of the first-order differential operators $\Xi=Vect(R^4)$. The
commutator of two operators  $\xi_1$ and $\xi_2$ gives the Lie bracket
formula for $\Xi$
\be
[\xi_1,\xi_2]=\xi^m_{br}\pa_m,\q\xi^m_{br}=\xi^n_1\pa_n\xi^m_2-\xi^n_2
\pa_n\xi^m_1.\lb{Liebr}
\ee
The finite diffeomorphisms belong to the universal enveloping algebra
$U\Xi$, for instance, the active diffeomorphism of the scalar field has
the following form:
\be
\phi^\prime(x)=(e^{-\cL_\xi}\phi)(x)=e^{-\xi}\phi e^\xi.
\ee

The gradient of the scalar field $\pa_m \phi$ transforms as the covariant
vector field
\be
\d_\xi(\pa_m\phi)=-(\xi\pa_m\phi)-(\pa_m\xi^p)\pa_p\phi=-\cL_\xi\pa_m\phi.
\ee
The contravariant vector field transforms as follows
\be
\d_\xi V^m=(-\xi+\pa_p\xi^m)V^p=-\cL_\xi V^m.
\ee
It is convenient to introduce the generators $L^p_q$ of the group GL(n,R)
\be
L^p_q T^{mr}_{ns}=\d^m_q T^{pr}_{ns}+\d^r_q T^{mp}_{ns}-\d^p_n T^{mr}_{qs}
-\d^p_s T^{mr}_{nq},\lb{Lgener}
\ee
then the compact form of the standard tensor transformation can be defined
via these generators and the multiplication of local parameters
$\xi^q_p(x)=\pa_p\xi^q$
\be
\d_\xi T^{mr}_{ns}=-\cL_\xi T^{mr}_{ns}=(-\xi+\xi^q_pL^p_q)T^{mr}_{ns}.
\ee

Let us consider the active transformation of the metric tensor
\bea
&&\d_\xi g_{mn}=-\cL_\xi g_{mn}=(-\xi+\xi^q_pL^p_q)g_{mn}.\lb{metrictr}
\eea
In the perturbation theory on the flat background metric $\eta_{mn}$, one
can analyze all nonlinearities in terms of the gauge gravitational field
$h_{mn}$
\bea
&&g_{mn}=\eta_{mn}+\k h_{mn},\nn\\
&&\d_\xi h_{mn}=-2\k^{-1}\xi_{(mn)}+(-\xi+\xi^q_pL^p_q)h_{mn},
\lb{hgaugetr}
\eea
where $\k$ is a gravitational constant and
\be
\xi_{(mn)}\equiv\sfrac12(\xi_{mn}+\xi_{nm}),\q\xi_{mn}=\eta_{np}\pa_m
\xi^p.
\ee
We  use also the equivalent representations of the  gauge field
$h^m_n=\eta^{mp}h_{pn},\q h^{mn}=\eta^{mp}\eta^{nq}h_{pq}$.

The inverse metric in this $h_{mn}$-representation is
\bea
g^{mn}=\eta^{mn}-\k h^{mn}+\k^2h^{mp}h^n_p-\k^3h^{mp}h^r_ph_r^n+\ldots
\eea
where all terms are covariant under the background Lorentz group. This
background invariance is useful, for instance, in the de Witt-Faddeev-Popov
quantization of gravity.

The basic transformation of the scalar field in the twisted diffeomorphism
group $U\Xi^{\cF}$  \cite{ABDMSW} has the following form:
\bea
\hd_\xi\phi=-\xi^m\pa_m\phi=-X^\star_\xi\star\phi=-\xi^m\star\pa_m\phi+
\sfrac{i}2\vt^{rs}(\pa_r\xi^m)\star\pa_s\pa_m\phi+O(\vt^2).
\eea
The alternative deformation of the diffeomorphism group $U\Xi_\star$
\cite{ADMW} uses the deformed Lie operator
\bea
\cL^\star_\xi\star\phi=\xi\star\phi=\xi^m\star\pa_m\phi
\eea
and  the deformed commutator
\bea
&&[\xi_1,\xi_2]_\star=\xi_1^m\pa_m\star\xi_2^n\pa_n-\xi_2^n\pa_n\star\xi_1^m\pa_m
+i\vt^{rs}(\pa_r\xi^n_2)\pa_n\star(\pa_s\xi^m_1)\pa_m+\ldots\nn\\
&&=[\xi_1,\xi_2]+\sfrac{i}2\vt^{rs}[[\pa_r,\xi_1],[\pa_s,\xi_2]]+\ldots
\eea
where the second line contains ordinary commutators of the first-order differential
operators. (The notation $[~,~]_\star$ should not be confused with the notation
$[~\stc~]$  for the $\star$-commutator \p{stcomm}.)
 The Hopf algebras $U\Xi^{\cF}$ and $U\Xi_\star$ are isomorphic by definition.
It was
shown that $U\Xi_\star$ transformations of the $\star$-product of functions
satisfy a more covariant form of the deformed Leibniz relation \cite{ADMW}.

In this work, we prefer to consider the undeformed $U\Xi^{\cF}$
transformations of the primary fields \p{scaltr},\p{metrictr}
\be
\hd_\xi\phi=-\xi\phi,\q\hd_\xi V_m=-\xi V_m-\pa_m\xi^nV_n
\ee
and the same Lie brackets \p{Liebr}. The coproduct in
$U\Xi^{\cF}$  is deformed by the twist operator $\cF$ \p{twist}
\bea
&&\D_t(\xi)=\exp(-\cP)(\xi\otimes 1+1\otimes \xi)\exp(\cP)=\xi\otimes 1
+1\otimes \xi\nn\\
&&-\sfrac{i}2\vt^{mn}([\pa_m,\xi]\otimes \pa_n+\pa_m\otimes[\pa_n,\xi])
\nn\\
&&-\sfrac18\vt^{mn}\vt^{rs}([\pa_m,[\pa_r,\xi]]\otimes \pa_n\pa_s
+\pa_m\pa_r\otimes[\pa_n,[\pa_s,\xi]])+O(\vt^3)
\eea
The corresponding twisted transformations of the noncommutative products
are defined via this coproduct, for instance,
\bea
&&\hd_\xi(\phi\star\chi)=-\mu_\star\circ\D_t(\xi)(\phi\otimes\chi)=
-(\xi(\phi\star\chi))=-X^\star_\xi\star(\phi\star\chi)
.
\eea
The last relation describes the decomposition of the $U\Xi^\cF$ transformation
in terms of the $U\Xi_\star$ operators including the simplest operator
\bea
&&-\cL^\star_\xi\star(\phi\star\chi)=-\xi\star(\phi\star\chi)\nn\\
&&=-(\xi\star\phi)\star\chi-\phi\star\xi\star\chi-i\vt^{rs}(\pa_r\phi)\star
\pa_s\xi^m\star\pa_m\chi+\ldots
\eea
satisfying the perfect generalization of the Leibniz rule. The corresponding
action of the $U\Xi_\star$ operator on vectors has the following form:
\be
-\cL^\star_\xi\star V_m=-\xi\star V_m-\pa_m\xi^n\star V_n.
\ee
 It is very important
for the noncommutative gravity that $\star$-products of scalar, vector or tensor
fields are the covariant objects both in $U\Xi^\cF$ and $U\Xi_\star$ groups.
For instance, the action invariant under $U\Xi^\cF$ transformations will be
automatically invariant with respect to $U\Xi_\star$.

We use the real symmetric metric tensor
$g_{mn}$ in the representation \p{hgaugetr}
as the basic field of the NC gravity.
In this treatment the deformed gravity describes the same number of the physical
degrees of freedom, as the gauge field of ordinary Einstein gravity.
The twist deformation guarantees the covariance of $\star$-products of
any primary tensors with respect to the transformations of $U\Xi^\cF$
\bea
&&
\hd_\xi(g_{mn}\star\phi)=(-\xi+\xi^q_pL^p_q)(g_{mn}\star\phi),\nn\\
&&\hd_\xi(g_{mn}\star\ldots\star g_{rs})=(-\xi+\xi^q_pL^p_q)
(g_{mn}\star\ldots\star g_{rs})\lb{transtar}
\eea
where the last formula contains an arbitrary number of fields. Note that
the operators $\pa_m$ and $L^p_q$ satisfy the undeformed Leibniz rules,
while the functions  $\xi^m(x)$ and $\xi^p_q(x)$ do not commute with
partial derivatives in the formula of the $\star$-product. The deformed
Leibniz rules for $\hd_\xi$ can be derived directly from these relations,
but we prefer not to use  these rules, because the covariance
relations \p{transtar} are more instructive.

\setcounter{equation}0
\section{Complex and real formalisms in the\\ noncommutative space}

 The covariant derivative of the vector $V_m$ in
the original geometric formalism of the noncommutative gravity \cite{ABDMSW}
formalism is defined via the left multiplication
\bea
\na_mV_n=\pa_mV_n-\hat\G^r_{mn}\star V_r ,\lb{leftcov}
\eea
where $\hat\G^r_{mn}$ is the left complex NC connection. The strict mathematical
analysis of this connection \cite{ADMW} shows that this geometric formalism
is a natural example of the quantum-group geometry. In particular, just in this
formalism one can formulate the simple generalization of the Leibniz rule for
the covariant derivative, for instance,
\be
 \na_m(V_n\star\phi) =(\na_mV_n)\star\phi+V_n\star\pa_m\phi,
\ee
although formulas for $\na_m(\phi\star V_n)$ or $\na_m(V_n\star V_r)$ are more
complicated. The corresponding complex NC Riemann tensor is
\bea
&&\hat{R}^{s}_{mnr}=\pa_m\hat\G^{s}_{nr}-\pa_n\hat\G^{s}_{mr}
+\hat\G^{p}_{nr}\star\hat\G^{s}_{mp}-
\hat\G^{p}_{mr}\star\hat\G^{s}_{np}.\lb{comcurv}
\eea
It is evident that the  covariant
derivative \p{leftcov} is complex for the real vector field.

Noncommutative products of real tensors or scalars are not real, in
general, but it is not difficult to construct the real  combinations in
algebra $A_\vt$, for instance, $\{g_{mn}\stc g_{rs}\}$. We  use
the reality condition in all constructions of the noncommutative geometry.

The noncommutative relation for the $\star$-inverse tensor $g_{mn}\star
G^{np}=\d^p_m$ was analyzed in ref.\cite{ABDMSW}. This
complex tensor  satisfies the conditions $\overline{G^{mn}}=G^{nm}\neq
G^{mn}$. The $\k h_{mn}$ decomposition of the $\star$-inverse tensor is
\bea
&&G^{mn}(h)=\eta^{mn}-\k h^{mn}+\k^2h^m_p\star h^{pn}
-\k^3h^m_p\star h^p_s\star h^{sn}+O(h^4).\lb{Ginv}
\eea
This formula can be used in the $\k h_{mn}$-decomposition of the complex
Christoffel symbols
\bea
\hat\G^p_{mn}(h)=\sfrac12\k(\pa_m h_{rn}+\pa_nh_{rm}-\pa_rh_{mn})\star G^{rp}(h).
\eea

The noncommutative gravity action \cite{ABDMSW} is real and its variation
with respect to the symmetric real tensor should yield directly the real tensor
NC equation of motion. Of course, one can consider the equivalent complex form
of this equation, however, the problem of reality is very important in the NC
gravity. Although this problem could be analyzed in the framework of the original
geometric formalism, we prefer to consider the manifestly real NC geometric
quantities at each stage of calculations.
Let us define, for instance, the real symmetric contravariant tensor
using the symmetric part of the complex tensor \p{Ginv}
\bea
&& g_\star^{mn}(h)=\sfrac12(G^{mn}+G^{nm})(h).\lb{gstar}
\eea
This tensor is not inverse to $g_{mn}$.
 To check transformation properties of the $\star$-perturbative
expressions, one should use the inhomogeneous transformations \p{hgaugetr}
which connect $\star$-polynomials of the  fields $h_{mn}$
\bea
&&\hd_\xi (h_{mn}\star h_{rs})=-2\k^{-1}\xi_{(mn)}h_{rs}
-2\k^{-1}\xi_{(rs)}h_{mn}-\cL_\xi(h_{mn}\star h_{rs}),\nn\\
&&\hd_\xi (h_{mn}\star h_{rs}\star h_{pq})=-2\k^{-1}\xi_{(mn)}
(h_{rs}\star h_{pq})-2\k^{-1}\xi_{(rs)}(h_{mn}\star h_{pq})\nn\\
&&-2\k^{-1}\xi_{(pq)}(h_{mn}\star h_{rs})
-\cL_\xi(h_{mn}\star h_{rs}\star h_{pq}).\lb{gaugetran}
\eea
These transformations are equivalent to the homogeneous transformations
of the tensor $\star$-polynomials \p{transtar}.

Note that  noncommutative generalizations of raising and lowering of
vector indices are not uniquely defined. One can consider, for instance,
the reality preserving relation between the  contravariant and covariant
vectors
\be
L_1:\q V^m\rightarrow V_m=\sfrac12\{g_{mn}\stc V^n\}=\eta_{mn}V^n+
\sfrac12\k\{h_{mn}\stc V^n\}.
\ee
The inverse map $L_1^{-1}$ is defined directly from this relation
\bea
V_m\rightarrow V^m=\eta^{mn}V_n-\sfrac12\k\{h^{mn}\stc V_n\}+\sfrac14
\k^2\{h^m_p\stc\{h^{pn}\stc V_n\}\}+\ldots\nn
\eea
An alternative raising procedure is also possible
$V_m\rightarrow \hat{V}^m=\sfrac12\{g^{mn}_\star\stc V_n\}$. It is not
difficult
to construct  reality-preserving raising or lowering maps for tensors.

The real NC determinant of the metric is defined via the 4th rank
antisymmetric symbols
\bea
&&g^\star\equiv D_4^\star(g_{mn})=\frac{1}{24}\ve^{m_1m_2m_3m_4}
\ve^{n_1n_2n_3n_4}g_{m_1n_1}\star g_{m_2n_2}\star g_{m_3n_3}\star
g_{m_4n_4}.\lb{gdens}
\eea
It transforms as the density of the weight -2 in $U\Xi^\cF$
\bea
&&\hd_\xi g^\star=-(\xi g^\star)-2\pa_p\xi^p g^\star.
\eea
The $\star$-polynomial $\k$-decomposition of this NC determinant is
\bea
&&g^\star\equiv -1-\k h^m_m-\sfrac12\k^2h^m_m\star h^n_n+\sfrac12\k^2
h^m_n\star h^n_m+O(h^3)=g(\k)+O_+(\vt^2),
\eea
where higher terms are omitted for brevity, and $g(k)$ is a determinant
of the classical metric. The absence of the odd-degree $\vt$-corrections
can be easily proved in $\star$-monomials of this decomposition.

The real NC generalization of the 4D volume density $e^\star$ can also be 
calculated as the perturbative $\star$-series
\bea
&&\hd_\xi e^\star=-(\xi e^\star)-\pa_p\xi^p e^\star,\nn\\
&& e^\star(\k,\vt)\equiv\sqrt{-g^\star}
=1+\sfrac12\k h^m_m+\sfrac18\k^2h^m_m\star h^n_n-\sfrac14\k^2h^m_n\star
h^n_m+O(h^3)\nn\\
&&=\sqrt{-g(\k)}+O_+(\vt^2).\lb{ncdens}
\eea

We shall use a subsidiary condition of reality of the NC connection
$\overline{\G^{\star r}_{mn}} =\G^{\star r}_{mn}$ which is compatible
with a standard definition of the gauge transformation of the connection
\bea
&&\hd_\xi \G^{\star r}_{mn}=-\pa_m\xi_n^r+(-\xi+\xi_p^qL^p_q)
\G^{\star r}_{mn}.
\eea
Left and right  products of the NC-connections and vectors obey 
similar inhomogeneous transformation laws
\bea
&&\hd_\xi(\G^{\star r}_{mn}\star V_t)=-(\pa_m\xi_n^r) V_t-\cL_\xi
(\G^{\star r}_{mn}\star V_t),\nn\\
&&\hd_\xi(V_t\star\G^{\star r}_{mn})=
-(\pa_m\xi_n^r) V_t-\cL_\xi(V_t\star\G^{\star r}_{mn}).
\eea

The symmetrized definition of the covariant $S$-derivatives is based on
anticommutators
\bea
&&\nabla_m^{S}\star V_n=\pa_m V_n-\sfrac12\{\G^{\star r}_{mn}\stc V_r\},
\nn\\
&&\nabla_m^S\star V^r=\pa_mV^r+\sfrac12\{\G^{\star r}_{pm}\stc V^p\},
\eea
it allows us to ensure reality and the correct tensor transformation
properties of these quantities. Note that the commutator term
$[\G^{\star r}_{mn}\stc V_r]$ is an independent $U\Xi^\cF$ tensor.
The covariant $S$-derivatives of tensors contain analogous anticommutator
terms with connections for any index. The real covariant NC derivative
does not satisfy the perfect generalization of the Leibniz rule, however,
it allows us to construct the manifestly real tensors of the NC gravity.

To obtain a real NC generalization of the Christoffel symbols, we use a
symmetrized version of the covariant constancy condition
\be
\pa_mg_{nr}=\sfrac12\{\G^{\star p}_{mn}\stc g_{pr}\}+\sfrac12
\{\G^{\star p}_{mr}\stc g_{np}\}.
\ee
For the symmetric connection $\G^{\star p}_{mn}=\G^{\star p}_{nm}$, this
condition yields the simple  relation
\bea
&&\k\g_{mnr}=\sfrac12\k(\pa_m h_{rn}+\pa_nh_{rm}-\pa_rh_{mn})=\eta_{rp}
\G^{\star p}_{mn}+\sfrac12\k \{h_{rp}\stc\G^{\star p}_{mn}\}
\eea
which can be solved by iterations. The $\star$-perturbative solution for
the real Christoffel symbols is
\bea
&&\G^{\star r}_{mn}=\k\eta^{rp}\g_{mnp}-\sfrac12\k^2 \{h^{rp}\stc
\g_{mnr}\}+\sfrac14\k^3\{h^r_s\stc\{ h^{sp}\stc\g_{mnp}\}\}\nn\\
&&-\sfrac{1}{8}\k^4\{h^r_t\stc\{h^t_s\stc\{ h^{sp}\stc\g_{mnp}\}\}\}+
O(h^5).\lb{mincon}
\eea
In Section 6 we analyze the alternative construction of the NC connection
which can be represented as the sum of our real connection plus some
tensor functions of the metric. In comparison with the classical
Christoffel symbols $\G^r_{mn}(\k)$ our minimal NC generalization has
only even-degree terms in the $\vt$-decomposition
\be
\G^{\star r}_{mn}(\k,\vt)=\G^r_{mn}(\k)+O_+(\vt^2)
\ee
The complex NC generalization of the Christoffel symbols \cite{ABDMSW}
contains terms odd in $\vt^{mn}$  including the linear term.

The trace of our connection
\bea
&&\G^{\star p}_{mp}=\sfrac12\k\pa_m h^r_r-\sfrac14\k^2\pa_m(h^{pr}\star
h_{pr})+\sfrac14\k^3A^3_m(h)+\ldots\nn
\eea
cannot be represented as a total derivative of some density starting
from the 3rd order term $A_m^3(h)$.

The real NC generalization of the Riemann tensor has the following form:
\bea
&&R^{\star s}_{mnr}=\pa_m\G^{\star s}_{nr}-\pa_n\G^{\star s}_{mr}
+\sfrac12\{\G^{\star p}_{nr}\stc\G^{\star s}_{mp}\}-
\sfrac12\{\G^{\star p}_{mr}\stc\G^{\star s}_{np}\}.
\eea
Note that the trace tensor $R^{\star r}_{mnr}=\pa_m\G^{\star r}_{nr}-
\pa_n\G^{\star r}_{mr}$ does not vanish in this representation. The
tensor transformation properties
\be
\hd_\xi R^{\star s}_{mnr}=(-\xi+\xi^p_qL^q_p)R^{\star s}_{mnr}
\ee
follow from the basic inhomogeneous transformations
\bea
&&\hd_\xi(\G^{\star p}_{nr}\star\G^{\star s}_{mt})=-(\pa_n\xi^p_r)
\G^{\star s}_{mt}-(\pa_m\xi^s_t)\G^{\star p}_{nr}-\cL_\xi
(\G^{\star p}_{nr}\star\G^{\star s}_{mt}).
\eea

The first Bianchi identity for this real NC Riemann tensor
\be
\ve^{upmr}R^{\star s}_{mnr}=0
\ee
is satisfied for the symmetric connection $\G^{\star s}_{mn}=
\G^{\star s}_{nm}$.
The deformed second Bianchi identity for $R^{\star s}_{mnr}$
\bea
&&\ve^{upmn}\na^S_pR^{\star s}_{mnr}=\sfrac12\ve^{upmn}(\{\G^{\star s}_{mz}
\stc\{\G^{\star t}_{pr}\stc\G^{\star z}_{nt}\}\}-\{\G^{\star t}_{pr}\stc
\{\G^{\star z}_{nt}\stc\G^{\star s}_{mz}\}\})\nn\\
&&=\sfrac12\ve^{upmn}[\G^{\star z}_{nt}\stc[\G^{\star t}_{pr}\stc
\G^{\star s}_{mz}]]
\eea
contains the unusual tensor function of connections vanishing in the
commutative limit. Note that an analogous Bianchi identity for the
complex NC Riemann tensor \p{comcurv} is also deformed.

In this formalism, the real NC Ricci tensor is symmetric  by  definition
\bea
&&R^\star_{(mr)}=\sfrac12(\pa_m\G^{\star n}_{rn}+\pa_r\G^{\star n}_{mn})
-\pa_n\G^{\star n}_{mr}+\sfrac12\{\G^{\star p}_{nr}\stc\G^{\star n}_{mp}\}
-\sfrac12\{\G^{\star p}_{mr}\stc\G^{\star n}_{pn}\}.
\eea
The contraction with $g^{mn}_\star$ \p{gstar} yields the corresponding
real scalar curvature
\be
R^\star=\sfrac12\{g^{mr}_\star\stc R^\star_{(mr)}\}=
\sfrac12\{g^{mr}_\star\stc R^{\star n}_{mnr}\}.
\ee
Thus, we consider the `minimal' formulation of the NC
geometry which preserves reality and gives only even-degree deformation
corrections to the classical geometric objects.

\setcounter{equation}0
\section{Minimal noncommutative gravity action and\\ equations}
The   real NC gravitational action can be constructed as a direct
analogue of the gravity action using our `minimal' formulation of the NC
geometry
\be
S_\star=\frac{1}{2\k^2}\int d^4x \{e^\star\stc R^\star\}=\frac{1}{\k^2}
\int d^4x e^\star\star R^\star,\lb{minimact}
\ee
where  $e^\star$ is the NC density \p{ncdens}. The NC-gravity-matter
action $A_\star=\int d^4x e^\star\star L_\star$ contains the real scalar
Lagrangian of additional matter fields $L_\star$. We treat $S_\star+
A_\star$ as the perturbation $\star$-series in $\k h_{mn}$.  In the next
Section we analyze the possible nonminimal versions of the NC-gravity
action using the pure noncommutative tensors and scalars vanishing in the
commutative limit.

To derive equations of motion in the noncommutative field theory one
should use the variation operator $\d_v$ satisfying the usual Leibniz
rule for $\star$-products
\be
\d_v (h_{mn}\star\ldots\star h_{pq})=(\d_v h_{mn}\star\ldots\star h_{pq})
+\ldots+ (h_{mn}\star\ldots\star\d_v h_{pq}),
\ee
where $\d_v h_{mn}$ are arbitrary infinitesimal tensor functions. Using
the variational principle and the cyclicity property of integrals on
$A_\vt$ one can obtain the NC gravitational equations
\bea
&&\d_v (S_\star+A_\star)=\frac{1}{\k^2}\int d^4x\,\d_vh_{mn}\star
\cE^{mn}=0\Rightarrow\nn\\
&&\cE^{mn}=\cG^{mn}(h)-\k^2{\cal T}^{mn}=0,
\eea
where contributions of all terms with $h_{mn}$ should be taken into
account. The contravariant real tensor density $\cE^{mn}$ contains the
pure gravitational part $\cG^{mn}$ and the NC generalization of the
matter energy-momentum tensor ${\cal T}^{mn}$. This tensor-density
equation can be transformed, for instance, to a more familiar covariant
tensor form of the NC equation using the NC procedure of lowering of the
tensor indices. Of course, the same real NC gravity equations can be
rewritten in terms of the complex NC  tensors of the original
geometric formalism.

The perturbative $\star$-expansion of the gravitational action
$S_\star(\k h)$ on the flat background contains the standard free spin 2
action $S^{(2)}(h)$ and the sum of higher-order interaction terms
$\k^{k-2}S^{(k)}_\star(h)$. The reality condition for $h_{mn}$ allows us
to preserve a standard number of the physical degrees of freedom in the
NC gravity.

The twisted gauge transformation of $h_{mn}$ \p{hgaugetr} connects terms
of different order. In addition, each term of the action is invariant
with respect to the background twisted Poincar\'e transformations with
constant parameters $c^r$ and $\o^{rs}$ considered in section 2
\bea
&&\hd_b (h_{mn}\star\ldots\star h_{pq})=[-c^r\pa_r+\o^{rs}(x_r\pa_s+
L_{rs})](h_{mn}\star\ldots\star h_{pq}),
\eea
where the last operator generates  Lorentz transformations of indices
$(\o^{rs}L_{rs})h_{mn}=-\o_{mr}h^r_n-\o_{nr}h^r_m$.

In our treatment, the minimal NC-gravity action and equations have only
even-degree $\vt$-corrections in comparison with the corresponding
classical action and equations
\bea
&&(S_\star+A_\star)(\k,\vt)=(S+A)(\k)+O_+(\vt^2),\nn\\
&& \cG^{mn}-\k^2\cT^{mn}=G^{mn}(\k)-\k^2T^{mn}
+O_+(\vt^2)=0,\lb{defequ}
\eea
where $G^{mn}(\k)-\k^2T^{mn}=0$ is the contravariant tensor-density
representation of the Einstein equations.

Let us consider the $\vt$-linear  approximate solution of this minimal
deformed gravity equations
\be
g_{mn}=g^{(0)}_{mn}(\k)+\sfrac12\vt^{rs}(\D g)_{mn,rs}(\k),
\lb{linpert}
\ee
where $g^{(0)}_{mn}(\k)$ is some classical gravity solution. The
$O_+(\vt^2)$ terms in Eq.\p{defequ} do not contribute to this
approximation, so $(\D g)_{mn,rs}$ can be found  from the undeformed
Einstein equations.

It is well known that the reality of the classical field action is
directly connected with the problem of unitarity in the corresponding
quantum theory. We do not analyze here the quantization problems, but
hope that noncommutativity could help solving some difficulties of the
quantum gravity.

\setcounter{equation}0
\section{ Pure noncommutative tensors in NC-gravity }

It should be remarked that the geometry of the noncommutative space based
on the twisted diffeomorphism group $U\Xi^\cF$ is much more flexible
than the Riemann geometry in the undeformed space. 
We have shown that the complex or real symmetric Christoffel symbols
and the curvature tensors can be constructed in terms of the same
real symmetric metric $g_{mn}$. Note that these alternative geometric objects
have the same commutative limits. The difference of alternative connections
is some pure noncommutative tensor with respect to $U\Xi^\cF$.
 These pure NC-tensors constructed from the basic
metric field can be independently added to the basic NC connection or
taken into account in the analysis of the NC gravity action in the framework
of any geometric formalism.  It is not clear a priori how to fix 
ambiguities arising from this NC flexibility, so
one should classify all pure NC tensor and scalar terms in this approach.

The  $\star$-commutator of the  gravitational fields $i[g_{mn}\stc
g_{rs}]$ is the simplest example of the pure NC tensor term. The
antisymmetric 2nd rank NC-tensor  can be expressed via the
complex inverse metric $G^{mn}$ \p{Ginv}
\bea
&&A^{mn}=\sfrac{i}2(G^{mn}-G^{nm})=\sfrac{i}2\k^2[h^m_p\stc h^{pn}]+O(\k^3).
\eea

It is not difficult to construct the pure NC-tensor of a length dimension
$l=-1$ , for instance, $[g_{rp}\stc \G^{\star p}_{mn}]$. The relation between
the complex NC Christoffel symbols \cite{ABDMSW} and our real symbols 
\p{mincon} contains just this tensor term
\bea
&&\hat\G^p_{mn}=\G^{\star p}_{mn}+\sfrac12[g_{tr}\stc\G^{\star r}_{mn}]\star
G^{tp}.
\eea
It is not also difficult to construct the inverse relation. Thus, all our
real NC tensors can be rewritten in terms of the complex connection
$\hat\G^p_{mn}$. Of course, the simple `minimal' objects of the real formalism
may contain a sum of different complex terms and vice versa...

In the NC formalism, an additional 
torsion-type tensor can be constructed from the $\star$-commutator
of the metric tensor and its derivatives 
\be
T_{[mn]}^{\star r}=\sfrac{i}2(\{g^{rp}_\star\stc[g_{ms}\stc
\G^{\star s}_{np}]\}-\{g^{rp}_\star\stc[g_{ms}\stc\G^{\star s}_{np}]\})
=O_-(\vt).
\ee
This pure NC tensor contains odd degrees of deformation constants, and it
could be used in the nonminimal NC gravity. 

One can also consider the following pure NC symmetric tensor  function
of dimension -2:
\bea
&&Q_{mn}=i[\G^{\star s}_{mn}\stc\G^{\star p}_{sp}]=-\vt^{ut}\pa_u
\G^s_{mn}(\k)\pa_t\G^p_{sp}(\k)+O_-(\vt^3)
\eea
and its scalar contraction with $g^{mn}_\star$.

Nonminimal versions of the NC gravity admit  additional odd
$\vt$-corrections to the minimal  action $S_\star+A_\star$. The
commutator terms in the quadratic action  are proportional to total
derivatives and do not deform the free spin 2 equations, however,
additional pure NC scalar terms could appear in the nonlinear gravity
interaction. Nonminimal NC gravity equations have the linear
$\vt$-corrections, and these terms can essentially change the deformed
gravity solutions. The $\star$-commutators of matter and gravitational
fields are important in the nonminimal NC action of the matter
interactions. For instance, the pure NC tensor $A^{[mn]}$  can
interact directly with the electromagnetic field-strength or
$\star$-commutators like $[\pa_m\phi\stc\pa_n\phi]$.

Note that the original action of the noncommutative gravity \cite{ABDMSW}
was constructed via the real part of the corresponding complex scalar
curvature $\hat{R}=G^{mn}\star \hat{R}^p_{mpn}$ in a tetrad
representation of the NC metric. It seems useful to rewrite their action
as the perturbation series in terms of $h_{mn}$ gauge field and compare it
with  our version of the action. It is clear that the lowest terms
of these actions are identical. Possible pure noncommutative tensors
 could contribute to higher-order terms in the $\k h_{mn}$ decomposition.

\setcounter{equation}0
\section{Conclusions}

In this paper, we addressed the problem of reality in the
$\vt^{mn}$-noncommutative gravity theory based on the twist-deformed
diffeomorphism group. The reality of the metric tensor is the important
physical condition and the deformed geometry formalism can preserve this
property. The real NC generalizations of the Christoffel symbols and Riemann
tensor are constructed in this approach. It is shown that the second Bianchi
identity for the NC Riemann tensor contains the pure noncommutative
tensor term. We considered the perturbative  expansions of all real NC
geometric quantities in terms of $\star$-polynomials of the gravitational field
$h_{mn}$ and its derivatives. These expansions are covariant under the 
background twisted Lorentz symmetry. The corresponding  real action of
the NC gravity and the real deformed gravity equations are presented.
We combine expansions of the geometric objects in the gravitational
constant $\k$ and the deformation constants, and control symmetry
properties of these perturbative methods. In our formulation,  these
NC-gravity equations are treated as minimal, because they have only 
even-degree $\vt$-corrections in comparison with the classical Einstein 
equations. The formalism of the
NC gravity based on the twisted diffeomorphism group is flexible: it
admits the existence of nonminimal pure noncommutative tensor and scalar
functions of $h_{mn}$ which have no classical analogues. It is not excluded 
that advantages of the NC gravity are connected just with the `nonminimal' 
odd-degree $\vt$-deformation  terms in the gravity interaction.
The reality-preserving flexible geometric formalism and the combined
$(\k,\vt)$ perturbation methods seem natural  for studying physical
effects of the noncommutative gravity theory, although  analogous effects
can also be  considered in the framework of the complex NC geometric formalism. 
We hope that arbitrary constants of the NC interactions could be fixed by
analyzing possible short-distance gravitational effects.

This work was partially supported by  DFG grant 436 RUS 113/669-3 , by
RFBR grants 06-02-16684 and 06-02-04012, by  NATO grant PST.GLG.980302
and by  grant of the Heisenberg-Landau programme.


\begin{thebibliography}{99}

\bibitem{CKNT}R. Oeckl, Nucl. Phys. {\bf B 581} (2000) 559, hep-th/0003018;
M. Chaichian, P.P. Kulish, K. Nishijima, A. Tureanu,
Phys. Lett. {\bf B 604} (2004) 98, hep-th/0408069.
\bibitem{We} J. Wess, Deformed coordinate spaces. Derivatives,
hep-th/0408080.
\bibitem{ABDMSW}P. Aschieri, C. Blohmann, M. Dimitrijevi\'c, F. Meyer,
P. Schupp, J. Wess, Class. Quant. Grav. {\bf 22} (2005) 3511,
hep-th/0504183.
\bibitem{ADMW}P. Aschieri,  M. Dimitrijevi\'c, F. Meyer, J. Wess,
Class. Quant. Grav. {\bf 23} (2006) 1883, hep-th/0510059.
\end{thebibliography}
\end{document}